\documentclass[aps,pra,draft,preprint,groupedaddress,showpacs]{revtex4}
\usepackage{epsfig}
\usepackage{amsmath}
\usepackage{amsfonts}

\def\bF{\mathbf F}
\def\bE{\mathbf E}
\def\bF{\mathbf F}
\def\bH{\mathbf H}
\def\bJ{\mathbf J}
\def\bB{\mathbf B}
\def\bD{\mathbf D}

\def\bI{\mathbf I}

\def\bx{\mathbf x}

\def\bk{\mathbf k}

\def\b0{\mathbf 0}
\def\bTheta{\mathbf \Theta}

\begin{document}

\title{Electromagnetic radiation in a time-varying background medium}
\author{Neil V. Budko}
\affiliation{Laboratory of Electromagnetic Research, Faculty of Electrical
Engineering, Mathematics and Computer Science,
Delft University of Technology,
Mekelweg 4, 2628 CD Delft, The Netherlands}
\email{n.v.budko@tudelft.nl}

\date{\today}

\begin{abstract}
Analytical solutions are presented for the electromagnetic radiation by an arbitrary pulsed source into a
homogeneous time-varying background medium. In the constant-impedance case an
explicit radiation formula is obtained for the synchronous permittivity and permeability described 
by any positive function of time. As might be expected, such a medium introduces significant 
spectral shifts and spatio-temporal modulation, which are analized here for the linear and 
exponential time-variations of the medium parameters. In the varying-impedance case the solution is obtained 
for the fourth-order polynomial time-dependence of the permittivity. In addition to the spectral shifts 
and modulation this spatially homogeneous medium scatters the field introducing causal echoes at the receiver location.
\end{abstract}


\maketitle

\section{Introduction}
One of the most difficult and least understood topics in science is the interaction between dynamic processes with separate physical origins and causes. Yet, such processes are very common in nature and engineering. For example, the electromagnetic field interacts with a non-autonomous (time-varying and independently controlled) medium whenever we try to visualize or measure an ultra-fast thermal, chemical or mechanical transition. A non-automous medium is also a good approximation for the nonlinear medium with a strong pumping field and a weak probing signal. Potential applications include: novel imaging algorithms for the visualization of ultra-fast dynamic processes in non-autonomous physical systems undergoing mechanical displacement, deformation, phase transition, chemical reaction, etc; detection of weak transient variations in the gravitational field; creation of ultra-short impulses and spatially localized light bullets; fast adaptive optical elements; broad-band extension of photonic and metamaterial crystals and associated effects, such as photonic band-gap waveguides and switches, perfect lenses, and invisibility cloaks; creation of congestion-free and transient-proof power, information, and transport networks. 

At present we are generally aware that the electromagnetic field evolution in a non-autonomous system cannot be reduced to a sequence of snapshots corresponding to the instantaneous values of the system parameters at various times. A plethora of very interesting and poorly understood transient effects is expected to be present. Some of those are: cross-modulation, velocity modulation, and parametric amplification of waves, e.g. dramatic pulse reshaping in attosecond optics, and exponential growth of disturbances in networks. So far, however, we do not have any general theoretical picture of these phenomena and do not understand connections between them, and therefore do not yet realize their full technological potential. Most of the practical modeling is numerical and does not allow for in-depth analysis. A few available, mainly one-dimensional, analytical models \cite{Shifrin,Nerukh,Bakunov} are often quasi-stationary, perturbative, restrict the type of the time-variation of the system parameters and the source, and have little to say about the near-field effects. Practically nothing is known about the influence of non-autonomous systems on the coherence of the field. Three-dimensional studies \cite{Shvartsburg, Tretyakov}
were mostly concerned with the closed systems, such as electromagnetic cavities and waveguides with perfectly conducting walls.
In general, the authors of non-perturbative studies agree that the time-variation of the medium at a scale comparable to the 
time-variation of the source results in a significant spatio-temporal modulation of the field.
Recently there has been some progress in the theory of media periodically varying in time \cite{Zurita-Sanchez}, where, as it turns out, one can apply the temporal analogue of the Bloch theorem. The analysis shows that such time-varying media introduce bandgaps in the wavenumber (momentum) domain, as opposed to the frequency-domain (energy) bandgaps produced by the spatially periodic media. 

To understand the spatio-temporal modulation of the time-domain source by the general time-varying medium one has to 
carry out a systematic theoretical analysis of the simplest non-autonomous electromagnetic system -- a homogeneous isotropic time-varying background medium. This is the subject of the present paper, where we shall consider the realistic vectorial three-dimensional field and arbitrary time-varying sources. This study is intended to fill some of the gaps in our knowledge, e.g. about the near-field behaviour, polarization effects, and a few-cycle pulsed sources. In addition, it provides valuable analytical soluitions, useful for the verification 
of numerical codes. The mathematical techniques applied here are standard, albeit involved. 
First, using a straightforward change of variable we obtain a completely explicit analytical solution for the 
constant-impedance case, where the time-variation of the permittivity and permeability is synchronous, but otherwise arbitrary 
(positive function of time). The more difficult time-varying impedance case is treated using an adaptation of the method originally 
proposed by Shvartsburg which allows to derive an explicit solution for a polynomial time-dependence of the permittivity. 

\section{The constant impedance case}
We begin with a relatively simple problem where the permittivity and pemeablity of the medium vary synchronously,
i.e.
\begin{align}
\label{eq:Synchronous}
\varepsilon(t)=\varepsilon_{0}a(t),\;\;\;\;\mu(t)=\mu_{0}a(t),
\end{align}
so that the medium impedance remains constant at all times, $\varepsilon(t)/\mu(t)=\varepsilon_{0}/\mu_{0}={\rm const}$.
Consider Maxwell's equations
 \begin{align}
 \label{eq:Maxwell}
 \begin{split}
 -\nabla\times \bH(\bx,t)+\partial_{t}\bD(\bx,t)&=-\bJ(\bx,t),
 \\
 \nabla\times\bE(\bx,t)+\partial_{t}\bB(\bx,t)&=0,
 \end{split}
 \end{align}
with the corresponding constitutive relations
 \begin{align}
 \label{eq:Constitutive}
 \bD(\bx,t)=\varepsilon_{0}a(t)\bE(\bx,t),\;\;\;\bB(\bx,t)=\mu_{0}a(t)\bH(\bx,t).
 \end{align}
Our goal is to obtain an explicit solution, called {\it radiation formula}, which gives the fields 
in terms of the given source current density $\bJ$.
Although, the procedure is pretty standard, it will be presented in full detail,
if only to clarify the differences with the case of time-varying impedance discussed in the 
next section. First, we rewrite the Maxwell equations as
 \begin{align}
 \label{eq:MaxwellDB}
 \begin{split}
 -\nabla\times \bB(\bx,t)+\mu_{0}a(t)\partial_{t}\bD(\bx,t)&=-\mu_{0}a(t)\bJ(\bx,t),
 \\
 \nabla\times\bD(\bx,t)+\varepsilon_{0}a(t)\partial_{t}\bB(\bx,t)&=0.
 \end{split}
 \end{align}
Now, we introduce a new variable with the physical dimension of time:
 \begin{align}
 \label{eq:VariableChange}
 \tau=\int\limits_{t_{0}}^{t}\frac{{\rm d}t'}{a(t')}.
 \end{align}
The original time $t$ will now be considered an implicit function $t(\tau)$ defined by 
the expression above. Taking into account that
 \begin{align}
 \label{eq:TauDerivative}
 \frac{\partial}{\partial t}=\frac{{\rm d} \tau}{{\rm d} t}\,\frac{\partial}{\partial \tau}
 =\frac{1}{a(t)}\frac{\partial}{\partial \tau}
 \end{align}
we rewrite the Maxwell equations in terms of variable $\tau$ as
 \begin{align}
 \label{eq:MaxwellTau}
 \begin{split}
 -\nabla\times \bB(\bx,\tau)+\mu_{0}\partial_{\tau}\bD(\bx,\tau)&=-\mu_{0}a(\tau)\bJ(\bx,\tau),
 \\
 \nabla\times\bD(\bx,\tau)+\varepsilon_{0}\partial_{\tau}\bB(\bx,\tau)&=0,
 \end{split}
 \end{align}
where the known functions $a(\tau)$ and $\bJ(\bx,\tau)$ must be understood as
$a(t(\tau))$ and $\bJ(\bx,t(\tau))$. The function $t(\tau)$ is the inverse mapping
with respect to the substiution (\ref{eq:VariableChange}). This inverse mapping 
exists and is one-to-one, if $a(t')>0$, $t_{0}\le t'\le t$. 
Maxwell's equations (\ref{eq:MaxwellTau}) are identical to the Maxwell equations in vacuum and
have a well-known analytical solution.
One usually starts by eliminating one of the unknowns, say $\bB$, and arriving at the following 
second-order vector wave equation for $\bD$:
 \begin{align}
 \label{eq:VectorWave}
 \nabla\times\nabla\times \bD(\bx,\tau)+\frac{1}{c_{0}^{2}}\partial^{2}_{\tau}\bD(\bx,\tau)=
 -\frac{1}{c_{0}^{2}}\partial_{\tau}\left[a(\tau)\bJ(\bx,\tau)\right],
 \end{align}
where $c_{0}=1/\sqrt{\varepsilon_{0}\mu_{0}}$. 
From (\ref{eq:MaxwellTau}) we derive the compatibility relation:
 \begin{align}
 \label{eq:Compatibility}
 \nabla\cdot\bD(\bx,\tau)=-\left(\nabla\cdot\,\right)\int\limits_{0}^{\tau}a(\tau')\bJ(\bx,\tau')\,{\rm d}\tau'.
 \end{align}
Rewriting (\ref{eq:VectorWave}) as
 \begin{align}
 \label{eq:VectorWaveDelta}
 \nabla\left(\nabla\cdot \bD(\bx,\tau)\right) - \Delta\bD(\bx,\tau) +\frac{1}{c_{0}^{2}}\partial^{2}_{\tau}\bD(\bx,\tau)=
 -\frac{1}{c_{0}^{2}}\partial_{\tau}\left[a(\tau)\bJ(\bx,\tau)\right],
 \end{align}
and applying (\ref{eq:Compatibility}), we arrive at
 \begin{align}
 \label{eq:ScalarWave}
 \Delta\bD(\bx,\tau) - \frac{1}{c_{0}^{2}}\partial^{2}_{\tau}\bD(\bx,\tau)=
 \frac{1}{c_{0}^{2}}\partial_{\tau}\bJ_{1}(\bx,\tau)
 -\nabla(\nabla\cdot\,)\int\limits_{0}^{\tau}\bJ_{1}(\bx,\tau')\,{\rm d}\tau',
 \end{align}
where we have introduced
 \begin{align}
 \label{eq:Current1}
 \bJ_{1}(\bx,\tau)&=a(\tau)\bJ(\bx,\tau),
 \end{align}
Performing the Laplace transform with respect to $\tau$ and the three-dimensional Fourier transform
with respect to $\bx$ we obtain the following algebraic equation in the $(\bk,s)$-domain
 \begin{align}
 \label{eq:ScalarWaveKS}
 -\vert\bk\vert^{2}\tilde{\bD}(\bk,s) - \gamma_{0}^{2}\tilde{\bD}(\bk,s)=
 \frac{s}{c_{0}^{2}}\tilde{\bJ}_{1}(\bk,s)
 +\bk(\bk\cdot\,)\frac{1}{s}\tilde{\bJ}_{1}(\bk,s),
 \end{align}
where $\gamma_{0}=s\sqrt{\varepsilon_{0}\mu_{0}}=s/c_{0}$. This equation can be solved as
 \begin{align}
 \label{eq:KSSolution}
 \tilde{\bD}(\bk,s)=
 -\frac{1}{\vert\bk\vert^{2}+\gamma_{0}^{2}}\left[\frac{s}{c_{0}^{2}}\tilde{\bJ}_{1}(\bk,s)
 +\bk(\bk\cdot\,)\frac{1}{s}\tilde{\bJ}_{1}(\bk,s)\right].
 \end{align}
Carrying out the inverse three-dimensional Fourier transform we obtain the $(\bx,s)$-domain solution as
 \begin{align}
 \label{eq:XSSolution}
 \hat{\bD}(\bx,s)=
 -\int\limits_{\bx'\in{\mathbb R}^{3}}g(\bx-\bx',s)\frac{s}{c_{0}^{2}}\hat{\bJ}_{1}(\bx',s)\,{\rm d}\bx'
 +\nabla(\nabla\cdot\,)\int\limits_{\bx'\in{\mathbb R}^{3}}g(\bx-\bx',s)
 \frac{1}{s}\hat{\bJ}_{1}(\bx',s)\,{\rm d}\bx',
 \end{align}
where the scalar Green's function is
 \begin{align}
 \label{eq:GreensFunction}
 g(\bx,s)=\frac{e^{-\gamma_{0}\vert\bx\vert}}{4\pi\vert\bx\vert}.
 \end{align} 
Computing the spatial derivatives and grouping the terms according to their spatial decay factors we get
 \begin{align}
 \label{eq:XSSpatialDerOut}
 \begin{split}
 \hat{\bD}(\bx,s)=&\int\limits_{\bx'\in{\mathbb R}^{3}}\frac{e^{-\gamma_{0}\vert\bx-\bx'\vert}}{4\pi\vert\bx-\bx'\vert^{3}}
 \left[3{\mathbb Q}-{\mathbb I}\right]\frac{1}{s}\hat{\bJ}_{1}(\bx',s)\,{\rm d}\bx'
 \\
 +&\int\limits_{\bx'\in{\mathbb R}^{3}}\frac{e^{-\gamma_{0}\vert\bx-\bx'\vert}}{4\pi\vert\bx-\bx'\vert^{2}}
 \left[3{\mathbb Q}-{\mathbb I}\right]\frac{1}{c_{0}}\hat{\bJ}_{1}(\bx',s)\,{\rm d}\bx'
 \\
 +&\int\limits_{\bx'\in{\mathbb R}^{3}}\frac{e^{-\gamma_{0}\vert\bx-\bx'\vert}}{4\pi\vert\bx-\bx'\vert}
 \left[{\mathbb Q}-{\mathbb I}\right]\frac{s}{c_{0}^{2}}\hat{\bJ}_{1}(\bx',s)\,{\rm d}\bx',
 \end{split}
 \end{align}
where ${\mathbb Q}=\bTheta(\bTheta\cdot\,)$, ${\bTheta}=(\bx-\bx')/\vert\bx-\bx'\vert$ and ${\mathbb I}$ is the Kronecker identity tensor.
The inverse Laplace transform of expressions (\ref{eq:XSSolution}) and (\ref{eq:XSSpatialDerOut}) gives 
the $(\bx,\tau)$-domain solution
 \begin{align}
 \label{eq:XTauEField}
 \begin{split}
 \bE(\bx,\tau)=&\int\limits_{\bx'\in{\mathbb R}^{3}}\frac{1}{4\pi\vert\bx-\bx'\vert^{3}}
 \left[3{\mathbb Q}-{\mathbb I}\right]
 \int\limits_{0}^{\tau_{\rm r}}\frac{a(\tau')\bJ(\bx',\tau')}{\varepsilon_{0}a(\tau)}\,{\rm d}\tau'\,{\rm d}\bx'
 \\
 +&\int\limits_{\bx'\in{\mathbb R}^{3}}\frac{1}{4\pi\vert\bx-\bx'\vert^{2}}
 \left[3{\mathbb Q}-{\mathbb I}\right]
 \frac{a(\tau_{\rm r})\bJ(\bx',\tau_{\rm r})}{c_{0}\varepsilon_{0}a(\tau)}\,{\rm d}\bx'
 \\
 +&\int\limits_{\bx'\in{\mathbb R}^{3}}\frac{1}{4\pi\vert\bx-\bx'\vert}
 \left[{\mathbb Q}-{\mathbb I}\right]
 \frac{\partial_{\tau}\left[a(\tau)\bJ(\bx',\tau)\right]_{\tau=\tau_{\rm r}}}{c_{0}^{2}\varepsilon_{0}a(\tau)}\,{\rm d}\bx',
 \end{split}
 \end{align}
Where we have used $\bD(\bx,\tau)=\varepsilon(\tau)\bE(\bx,\tau)$, and 
(\ref{eq:Current1}). The retarded $\tau$-time is
 \begin{align}
 \label{eq:RetardedTau}
 \tau_{\rm r}=\tau-\frac{\vert\bx-\bx'\vert}{c_{0}}.
 \end{align}
To obtain the $(\bx,t)$-domain solution we now need to introduce the inverse $\tau$-function $t(\tau)$, which assignes a particular and unique proper time $t$, to any given $\tau$. If the medium is time-invariant, then $t(\tau)=t$, $\mu(\tau)=\mu_{0}$, $\varepsilon(\tau)=\varepsilon_{0}$, reducing (\ref{eq:XTauEField}) to the well-known free-space radiation formula (see e.g. \cite{Budko2009a}): 
 \begin{align}
 \label{eq:XTTimeInvariant}
 \begin{split}
 \bE(\bx,t)=&\int\limits_{\bx'\in{\mathbb R}^{3}}\frac{1}{4\pi\vert\bx-\bx'\vert^{3}}
 \left[3{\mathbb Q}-{\mathbb I}\right]
 \int\limits_{t_{0}}^{t_{\rm r}}\frac{\bJ(\bx',t')}{\varepsilon_{0}}\,{\rm d}t'\,{\rm d}\bx'
 \\
 +&\int\limits_{\bx'\in{\mathbb R}^{3}}\frac{1}{4\pi\vert\bx-\bx'\vert^{2}}
 \left[3{\mathbb Q}-{\mathbb I}\right]
 \frac{\bJ(\bx',t_{\rm r})}{c_{0}\varepsilon_{0}}\,{\rm d}\bx'
 \\
 +&\int\limits_{\bx'\in{\mathbb R}^{3}}\frac{1}{4\pi\vert\bx-\bx'\vert}
 \left[{\mathbb Q}-{\mathbb I}\right]
 \frac{\left[\partial_{t}\bJ(\bx',t)\right]_{t=t_{\rm r}}}{c_{0}^{2}\varepsilon_{0}}\,{\rm d}\bx',
 \end{split}
 \end{align}
and the retarded time is $t_{\rm r}=t-\vert\bx-\bx'\vert/c_{0}$. Comparing (\ref{eq:XTTimeInvariant}) and (\ref{eq:XTauEField})
we see that the difference between the time-varying and time-invariant (constant-impedance)
background media is in the presence of the time-weghting and in the modulation of the source current density.
It is interesting to note that although the formally defined velocity of light in a time-varying medium 
would be also time-varying, i.e. $c=1/\sqrt{\varepsilon(t)\mu(t)}=c_{0}/a(t)$, the radiation formula (\ref{eq:XTauEField}) features 
the time delay factor $\vert\bx-\bx'\vert/c_{0}$, see (\ref{eq:RetardedTau}), containing the usual 
constant speed of light in vacuum. In other words, the propagation of light in a time-varying constant-impedance
medium may be viewed as a propagation in vacuum where the causal time-delay is measured in terms 
of the ``universal'' $\tau$-time. The actual measurements however are done in terms of the usual $t$-time.
Hence, to translate the abstract $\tau$-domain picture into the $t$-domain reality, we need to find the inverse 
mapping $t(\tau)$ and use it in the given functions $a(t)$ and $\bJ(\bx,t)$ thus obtaining the 
functions $a(\tau)$ and $\bJ(\bx,\tau)$ appearing in the radiation formula (\ref{eq:XTauEField}). 
Then, after performing all the necessary mathematical operations with respect the $\tau$-variable 
in the right-hand-side of (\ref{eq:XTauEField}), we shall try to recognize the $t$-domain solution.

Although, the inverse function $t(\tau)$ exists for any $a(t')>0$, $t_{0}\le t' \le t$, it can be found explicitly
only if the integral in (\ref{eq:VariableChange}) can be evaluated analytically and the subsequent algebraic equation can be 
analytically solved for $t$. Consider a linear time-dependence:
 \begin{align}
 \label{eq:LinearA}
 a(t)=a_{0}+bt.
 \end{align}
Then, from (\ref{eq:VariableChange}) we find
 \begin{align}
 \label{eq:LinearTTau}
 t(\tau)=\frac{a_{0}+bt_{0}}{b}e^{b\tau}-\frac{a_{0}}{b},
 \end{align}
and 
 \begin{align}
 \label{eq:LinearATTau}
 a(\tau)=(a_{0} + bt_{0})e^{b\tau}.
 \end{align}
A time-harmonic source $\bJ(\bx,t)=\bI(\bx)\cos(\omega t)$ will have the following representation in the $\tau$-domain
 \begin{align}
 \label{eq:HarmonicCurrent}
 \bJ(\bx,\tau)=\bI(\bx)\cos(\omega t(\tau))=\bI(\bx)\cos\left[\omega\left(\frac{a_{0}+bt_{0}}{b}e^{b\tau}-\frac{a_{0}}{b}\right)\right].
 \end{align}
In the far-field zone the electric field strength is proportional to the last term in (\ref{eq:XTauEField}). Explicitly 
we obtain
 \begin{align}
 \label{eq:FarFieldEHarmonic}
 \begin{split}
 \bE(\bx,\tau)&\approx
 \int\limits_{\bx'\in{\mathbb R}^{3}}
 \frac{\left[{\mathbb Q}-{\mathbb I}\right]\bI(\bx')
 \partial_{\tau}\left\{(a_{0} + bt_{0})e^{b\tau}
 \cos\left[\omega\left(\frac{a_{0}+bt_{0}}{b}e^{b\tau}-\frac{a_{0}}{b}\right)\right]\right\}_{\tau=\tau_{\rm r}}}
 {4\pi c_{0}^{2}\varepsilon_{0}\vert\bx-\bx'\vert(a_{0} + bt_{0})e^{b\tau}}\,{\rm d}\bx'
  \\ 
 &= \int\limits_{\bx'\in{\mathbb R}^{3}}
 \frac{\left[{\mathbb Q}-{\mathbb I}\right]\bI(\bx')
 b e^{b\tau_{\rm r}} \cos\left[\omega\left(\frac{a_{0}+bt_{0}}{b}e^{b\tau_{\rm r}}-\frac{a_{0}}{b}\right)\right]}
 {4\pi c_{0}^{2}\varepsilon_{0}\vert\bx-\bx'\vert e^{b\tau}}\,{\rm d}\bx'
  \\ 
 &+  \int\limits_{\bx'\in{\mathbb R}^{3}}
  \frac{\left[{\mathbb Q}-{\mathbb I}\right]\bI(\bx')
 \omega (a_{0}+bt_{0}) e^{2b\tau_{\rm r}} \sin\left[\omega\left(\frac{a_{0}+bt_{0}}{b}e^{b\tau_{\rm r}}-\frac{a_{0}}{b}\right)\right]}
 {4\pi c_{0}^{2}\varepsilon_{0}\vert\bx-\bx'\vert e^{b\tau}}\,{\rm d}\bx'
 \\ 
 &= \int\limits_{\bx'\in{\mathbb R}^{3}}
 \frac{\left[{\mathbb Q}-{\mathbb I}\right]\bI(\bx')
 b e^{-\frac{b}{c_{0}}\vert\bx-\bx'\vert} \cos\left[\omega \left(\frac{a_{0}+bt_{0}}{b}e^{b\tau} e^{-\frac{b}{c_{0}}\vert\bx-\bx'\vert}-\frac{a_{0}}{b}\right)\right]}
 {4\pi c_{0}^{2}\varepsilon_{0}\vert\bx-\bx'\vert }\,{\rm d}\bx'
  \\ 
 &+  \int\limits_{\bx'\in{\mathbb R}^{3}}
  \frac{\left[{\mathbb Q}-{\mathbb I}\right]\bI(\bx')
 \omega (a_{0}+bt_{0}) e^{b\tau} e^{-2\frac{b}{c_{0}}\vert\bx-\bx'\vert}
 \sin\left[\omega \left(\frac{a_{0}+bt_{0}}{b}e^{b\tau}e^{-\frac{b}{c_{0}}\vert\bx-\bx'\vert}-\frac{a_{0}}{b}\right)\right]}
 {4\pi c_{0}^{2}\varepsilon_{0}\vert\bx-\bx'\vert}\,{\rm d}\bx'
 \end{split} 
 \end{align}
Where, using (\ref{eq:LinearTTau}) and (\ref{eq:LinearATTau}), we recognize the $t$-domain result as
 \begin{align} 
 \label{eq:FarFieldEHarmonicTdomain}
 \begin{split}
 \bE(\bx,t)&\approx
 \int\limits_{\bx'\in{\mathbb R}^{3}}
 \frac{\left[{\mathbb Q}-{\mathbb I}\right]\bI(\bx')
 b e^{-\frac{b}{c_{0}}\vert\bx-\bx'\vert} \cos\left(\omega' t+\frac{a_{0}}{b}(\omega'-\omega)\right)}
 {4\pi c_{0}^{2}\varepsilon_{0}\vert\bx-\bx'\vert }\,{\rm d}\bx'
  \\ 
 &+  \int\limits_{\bx'\in{\mathbb R}^{3}}
  \frac{\left[{\mathbb Q}-{\mathbb I}\right]\bI(\bx')
 \omega' a(t) e^{-\frac{b}{c_{0}}\vert\bx-\bx'\vert} \sin\left(\omega' t+\frac{a_{0}}{b}(\omega'-\omega)\right)}
 {4\pi c_{0}^{2}\varepsilon_{0}\vert\bx-\bx'\vert}\,{\rm d}\bx'.
 \end{split} 
 \end{align}
In this formula we have a new frequency
 \begin{align}
 \label{eq:FrequencyShift}
 \omega'=\omega e^{-\frac{b}{c_{0}}\vert\bx-\bx'\vert},
 \end{align}
showing that one of the effects due to the time-varying background is a shift of the observed frequency depending on the
distance between the source and the receiver. In particular, a medium varying linearly with time at a rate $b$ 
introduces a frequency shift varying exponentially with distance (\ref{eq:FrequencyShift}) -- 
blueshift for negative $b$, red-shift for positive. We also notice an additional exponential distance 
factor in (\ref{eq:FarFieldEHarmonicTdomain}) -- decay for redshift, amplification for blueshift. 
Tracing back to (\ref{eq:XTauEField}) we see that the latter parametric amplification is due to the modulation of the source by 
the $a(\tau)$-function, which produces other interesting phenomena as well. For example, if the time-variation 
of the current is such that
 \begin{align}
 \label{eq:BlackOut}
 \partial_{\tau}\left[a(\tau)\bJ(\bx,\tau)\right]=0,\;\;\;\text{i.e.}\;\;\;a(\tau)\bJ(\bx,\tau)=\,{\rm const},
 \end{align}
then the far-field term in (\ref{eq:XTauEField}) disappears, meaning that the field will die-off very rapidly with 
the distance from the source, and that the time-varying current, which would normally radiate, may be effectively
silenced by a suitably time-varying homogeneous and lossless medium.

The other related but opposite phenomenon is the radiation by a DC-like current. Normally, a constant current
does not radiate (into the far-field zone), neither does a charged particle moving at a constant velocity.
Whereas, in a time-varying medium, a constant current would produce a far-field radiation proportional to the 
magnitude of the current and with the time-characteristics depending on $a(t)$. Indeed,
 \begin{align}
 \label{eq:DCRadiation}
 \partial_{\tau}\left[a(\tau)\bJ(\bx)\right]=\bJ(\bx)\partial_{\tau}a(\tau)\ne 0.
 \end{align}
Obviously, it is the medium, which is the true cause of radiation in this case. 

All of this confirms that a 
time-varying medium, even a spatially homogeneous one, may produce a significant space-time modulation of 
the electromagnetic field. Although, a medium with constant impedance considered so far is a very rare thing,
at least one physically interesting instant of such a medium is found in General Relativity Theory (GRT) \cite{Leonhardt, Budko2009b},
where a linearly expanding universe corresponds to the linear $a(t)$ discussed above (apart for the
current-modulating factor, which is $a^{4}(\tau)$ in GRT). In particular, application of the present analysis
to the problem of Hubble's redshift law shows that a linearly expanding universe produces an exponential
redshift, not a linear one as it is currently believed. In general, it may be shown, using only (\ref{eq:LinearTTau}) 
and (\ref{eq:RetardedTau}), that any two temporal features 
of the time-derivative of the modulated current density, say the zeros or extrema of the signal, radiated by the source at times 
$t_{1}^{\rm src}$ and $t_{2}^{\rm src}$, with an interval $\Delta t^{\rm src}=t_{1}^{\rm src}-t_{2}^{\rm src}$ between them,
will be detected by the receiver at times $t_{1}^{\rm rec}$ and $t_{2}^{\rm rec}$ with a different 
interval $\Delta t^{\rm rec}=t_{1}^{\rm rec}-t_{2}^{\rm rec}$, namely:
 \begin{align}
 \label{eq:IntervalLinear}
 \frac{\Delta t^{\rm rec}}{\Delta t^{\rm src}}=e^{b\vert\bx^{\rm src}-\bx^{\rm rec}\vert},
 \end{align}
where $\bx^{\rm src}$ and $\bx^{\rm rec}$ are the spatial locations of the source and the receiver, respectively.
This is simply a generalization of the frequency shift law (\ref{eq:FrequencyShift}). In combination with the radiation
formula (\ref{eq:XTauEField}) and the time-delay expression (\ref{eq:RetardedTau}) this law provides an intuitively
appealing picture of the expanding universe. First of all, there is no relative motion of the source and receiver, as the position
vectors in the final radiation formula (\ref{eq:XTauEField}) are time-independent. The speed of light also 
does not change as far as the $\tau$-domain picture is concerned. Yet, in the real $t$-domain world we observe 
the stretching of the time-interval between certain characteristic events in the received signal according 
to (\ref{eq:IntervalLinear}). 

Imagine two clocks, which are intially at the same location in space and are synchronized. Suppose that each second
these clocks radiate a short impulse of light. Now, we place one of these clocks some distance away and observe that the interval 
between the impulses arriving from that clock is longer than one second (impulses have become broader in time too).
We are naturally inclined to think that the remote clock started to walk slower for some reason, and decide to go and check it.
To our surprise, however, upon our arrival we find that everything is fine with this remote clock, and it is the other one which 
walks slower, judging from the impulses that are coming from it. We begin to suspect that there is a redshift involved 
due to the relative motion of the two clocks. Yet, when we put a ruler between them and observe the measured distance
for some time, nothing seems to happen, and the clocks appear to be at rest with respect to eachother. This, in short, 
is what happens in an expanding universe. 

Due to its importance in cosmology, especially for understanding the process of inflation proposed within the Big Bang theory, 
we shall consider here the case of exponentially growing $a(t)$ as well. Suppose,
 \begin{align}
 \label{eq:AExp}
 a(t)=a_{0}e^{bt}.
 \end{align}
Then, from (\ref{eq:VariableChange}), the inverse function $t(\tau)$ is found to be
 \begin{align}
 \label{eq:TTauExp}
 t(\tau)=-\frac{1}{b}\ln\left[e^{-bt_{0}}-a_{0}b\tau\right].
 \end{align}
It is obvious that this inverse mapping is only defined for $\tau < e^{-bt_{0}}/(a_{0}b)$, since otherwise
the physical time $t$ would become complex. Therefore, according
to (\ref{eq:RetardedTau}), we may conclude that the light impulse radiated at $\tau$ will be confined to the spatial radius
 \begin{align}
 \label{eq:FieldUndefined}
 R=\frac{c_{0}}{a_{0}b}e^{-bt_{0}}-c_{0}\tau,
 \end{align}
effectively excluding any communication between points further than $R$ apart. 
Alternatively, one could speculate that the laws of light propagation are different beyond that radius, which is highly unlikely.
Computing the ratio of time intervals between the corresponding features (e.g. zeros) of the electromagnetic signal
at the source and receiver locations we get the following expression
 \begin{align}
 \label{eq:LogShift}
 \frac{\Delta t ^{\rm rec}}{\Delta t ^{\rm src}}=
 \log_{s}r,
 \end{align}
where
 \begin{align}
 \label{eq:S}
 \begin{split}
 s&=\frac{e^{-bt_{0}}-a_{0}b\tau_{2}^{\rm src}}{e^{-bt_{0}}-a_{0}b\tau_{1}^{\rm src}},
 \\
 r&=\frac{e^{-bt_{0}}-a_{0}b\left(\tau_{2}^{\rm src}+\vert\bx-\bx'\vert/c_{0}\right)}
 {e^{-bt_{0}}-a_{0}b\left(\tau_{1}^{\rm src}+\vert\bx-\bx'\vert/c_{0}\right)}.
 \end{split}
 \end{align}
This proves wrong the common assumption that an exponentially expanding universe introduces
an exponential cosmological redshift. The redshift appears also to depend upon the interval between
$\tau_{1}^{\rm src}$ and $\tau_{2}^{\rm src}$, i.e. is inhomogeneous -- different frequencies will
have different shifts.
\section{Time-varying impedance}
The problem for a general time-varying impedance with asynchronously varying permittivity and permeability 
so far resists analytical treatment. However, for a class of time-varying impedances, 
where $\mu(t)$ is a constant, and $\varepsilon(t)$ is a fourth-order polynomial in time, 
a useful analytical result can still be obtained. The solution can even be put in a form of a radiation formula similar to 
(\ref{eq:XTTimeInvariant}) and (\ref{eq:XTauEField}) revealing important differences with the stationary medium 
and the constant-impedance cases.
The technique is due to A.~B.~Shvartzburg, who considered mainly the source-free one-dimensional Maxwell's 
equations. Here this method is extended to arbitrary pulsed sources in three dimensions. 
We denote:
 \begin{align}
 \label{eq:EpsilonU}
 \varepsilon(t)=\varepsilon_{0}a(t)=\varepsilon_{0}u^{2}(t),\;\;\;\mu(t)=\mu_{0}={\rm const}.
 \end{align}
We again introduce a new (slightly different) time-like variable:
 \begin{align}
 \label{eq:DefTau}
 \tau=\int\limits_{t_{0}}^{t}\frac{{\rm d}t'}{\sqrt{a(t')}}=\int\limits_{t_{0}}^{t}\frac{{\rm d}t'}{u(t')},
 \end{align}
so that the partial derivatives are related as
 \begin{align}
 \label{eq:PartialTau}
 \frac{\partial}{\partial t}=\frac{1}{u(t)}\frac{\partial}{\partial{\tau}}.
 \end{align}
In addition to this variable change, a new unknown function is introduced as
 \begin{align}
 \label{eq:DefF}
 \bF(\bx,t)=\frac{\bD(\bx,t)}{\sqrt{u(t)}}.
 \end{align}
To deal with the three-dimensional vector case some preliminary work needs to be done, consiting of deriving 
the $(\bx,\tau)$-domain compatibility relation for the $\bF$-field. It can be obtained from the $(\bx,\tau)$-domain 
Maxwell's equations, and turns out to be
 \begin{align}
 \label{eq:FCompatibility}
 \nabla\cdot\bF(\bx,\tau)=-\frac{1}{\sqrt{u(\tau)}}(\nabla\cdot)\int\limits_{0}^{\tau}u(\tau')\bJ(\bx,\tau')\,{\rm d}\tau'.
 \end{align}
Now we go back to the $(\bx,t)$-domain Maxwell's equations and eliminate $\bH$, thus obtaining the second order 
vector wave equation for $\bD$:
 \begin{align}
 \label{eq:Scalarwave}
 \nabla\times\nabla\times\bD(\bx,t)+\frac{u^{2}(t)}{c_{0}^{2}}\partial_{t}^{2}\bD(\bx,t)=-\frac{u^{2}(t)}{c_{0}^{2}}\partial_{t}\bJ(\bx,t).
 \end{align}
Performing both the function and the variable change, and using (\ref{eq:FCompatibility}) we arrive at
\begin{align}
\label{eq:Substitution}
\begin{split}
&\Delta\bF(\bx,\tau)-\frac{1}{c_{0}^{2}}\partial_{\tau}^{2}\bF(\bx,\tau)
-\frac{1}{c_{0}^{2}}\left[\frac{1}{2}u(t)\partial_{t}^{2}u(t)-\frac{1}{4}\left(\partial_{t}u(t)\right)^{2}\right]\bF(\bx,\tau)
\\
&=\frac{\sqrt{u(\tau)}}{c_{0}^{2}}\partial_{\tau}\bJ(\bx,\tau) - \frac{1}{\sqrt{u(\tau)}}
\nabla(\nabla\cdot\,)\int\limits_{0}^{\tau}u(\tau')\bJ(\bx,\tau')\,{\rm d}\tau'.
\end{split}
\end{align}
Now, and this is the essence of the Shwartzburg method, we shall limit the time-dependence of the medium, i.e. the class
of functions $u(t)$, to those satisfying the nonlinear equation
\begin{align}
\label{eq:Uequation}
\frac{1}{2}u(t)\partial_{t}^{2}u(t)-\frac{1}{4}\left(\partial_{t}u(t)\right)^{2}=\frac{1}{T^{2}},
\end{align}
where $T$ is a real constant with the physical dimension of time. In that case (\ref{eq:Scalarwave}) 
reduces to
\begin{align}
\label{eq:Tauproblem}
\begin{split}
&\Delta\bF(\bx,\tau)-\frac{1}{c_{0}^{2}}\partial_{\tau}^{2}\bF(\bx,\tau)
-\frac{1}{T^{2}c_{0}^{2}}\bF(\bx,\tau)
\\
&= \frac{\sqrt{u(\tau)}}{c_{0}^{2}}\partial_{\tau}\bJ(\bx,\tau)
 -\frac{1}{\sqrt{u(\tau)}}\nabla(\nabla\cdot\,)\int\limits_{0}^{\tau}u(\tau')\bJ(\bx,\tau')\,{\rm d}\tau',
\end{split}
\end{align}
which can be solved analytically for any source.
The functions $u(t)$ leading to this simplified equation are the solutions of (\ref{eq:Uequation}) and 
have the form:
\begin{align}
\label{eq:Utype}
u(t)=a+bt+ct^{2}.
\end{align}
This polynomial function allows to model both rising and decaying profiles of $\varepsilon(t)=\varepsilon_{0}u^{2}(t)$, 
while parameter $T$ determines the temporal scale of variations.

The solution of (\ref{eq:Tauproblem}) is obtained via the same path as in the previous section.
First, we transform the equation to the $(\bk,s)$-domain and solve it there. Then, we carry out the inverse
Fourier transform and obtain the following $(\bx,s)$-domain solution:
 \begin{align}
 \label{eq:XSFsolution}
 \begin{split}
 \hat{\bF}(\bx,s)=&\int\limits_{\bx'\in{\mathbb R}^{3}}
 \frac{e^{-\gamma\vert\bx-\bx'\vert}}{4\pi\vert\bx-\bx'\vert^{3}}
 \left[3{\mathbb Q}-{\mathbb I}\right]\hat{\bJ}_{1}(\bx',s)\,{\rm d}\bx'
 \\
 &+\int\limits_{\bx'\in{\mathbb R}^{3}}
 \frac{\gamma e^{-\gamma\vert\bx-\bx'\vert}}{4\pi\vert\bx-\bx'\vert^{2}}
 \left[3{\mathbb Q}-{\mathbb I}\right]\hat{\bJ}_{1}(\bx',s)\,{\rm d}\bx'
 \\
 &+\int\limits_{\bx'\in{\mathbb R}^{3}}
 \frac{\gamma^{2} e^{-\gamma\vert\bx-\bx'\vert}}{4\pi\vert\bx-\bx'\vert}
 {\mathbb Q}\hat{\bJ}_{1}(\bx',s)\,{\rm d}\bx'
 \\
 &-\int\limits_{\bx'\in{\mathbb R}^{3}}
 \frac{e^{-\gamma\vert\bx-\bx'\vert}}{4\pi\vert\bx-\bx'\vert}
 \frac{1}{c_{0}^{2}}\hat{\bJ}_{2}(\bx',s)\,{\rm d}\bx',
 \end{split} 
 \end{align}
where the currents are the $(\bx,s)$-domain images of
 \begin{align}
 \label{eq:XTauCurents}
 \begin{split}
 \bJ_{1}(\bx,\tau)=&\frac{1}{\sqrt u(\tau)}\int\limits_{0}^{\tau}u(\tau')\bJ(\bx,\tau')\,{\rm d}\tau',
 \\
 \bJ_{2}(\bx,\tau)=&\sqrt{u(\tau)}\partial_{\tau}\bJ(\bx,\tau).
 \end{split}
 \end{align}
The difference with (\ref{eq:XSSpatialDerOut}) is mainly in the mathematical form of the parameter $\gamma$,
which is now
 \begin{align}
 \label{eq:NewGamma}
 \gamma=\frac{1}{c_{0}}\sqrt{s^{2}+1/T^{2}}.
 \end{align}
Obviously, for $1/T^{2}=0$, corresponding to the infinitely slow variation of $\varepsilon(t)$, we recover the free-space case
$\gamma=s/c_{0}$.
To transform the result to the $(\bx,\tau)$-domain, we recall the following known Laplace transform pairs:
 \begin{align}
 \label{eq:Abr1}
 e^{-k\sqrt{s^{2}+a^{2}}},\;\;\; (k>0) \;\;\;&\rightarrow \;\;\;\delta(t-k)-\frac{ak}{\sqrt{t^{2}-k^{2}}}J_{1}(a\sqrt{t^{2}-k^{2}})H(t-k),
 \\
 \label{eq:Abr2}
 \frac{e^{-k\sqrt{s^{2}+a^{2}}}}{\sqrt{s^{2}+a^{2}}},\;\;\; (k\ge 0) \;\;\;&\rightarrow \;\;\;J_{0}(a\sqrt{t^{2}-k^{2}})H(t-k),
 \end{align}
where $H(t-k)$ is a unit step function located at $t=k$. Using these transforms we can write the result as
 \begin{align}
 \label{eq:XTauFsolution}
 \begin{split}
 \bF(\bx,\tau)=&\int\limits_{\bx'\in{\mathbb R}^{3}}
 \frac{1}{4\pi\vert\bx-\bx'\vert^{3}}
 \left[3{\mathbb Q}-{\mathbb I}\right]\bJ_{1}(\bx',\tau_{\rm r})\,{\rm d}\bx'
 \\
 &-\frac{1}{c_{0}T}\int\limits_{\bx'\in{\mathbb R}^{3}}
 \frac{1}{4\pi\vert\bx-\bx'\vert^{2}}
 \left[3{\mathbb Q}-{\mathbb I}\right]\int\limits_{0}^{\tau}
 g_{1}(\bx-\bx',\tau-\tau')\bJ_{1}(\bx',\tau')\,{\rm d}\tau'\,{\rm d}\bx'
 \\
 &+\frac{1}{c_{0}}\int\limits_{\bx'\in{\mathbb R}^{3}}
 \frac{1}{4\pi\vert\bx-\bx'\vert^{2}}
 \left[3{\mathbb Q}-{\mathbb I}\right]
 \int\limits_{0}^{\tau}
 g_{0}(\bx-\bx',\tau-\tau')
 \left[\partial_{\tau'}^{2}+\frac{1}{T^{2}}\right]\bJ_{1}(\bx',\tau')\,{\rm d}\tau'\,{\rm d}\bx'
 \\
 &+\frac{1}{c_{0}^{2}}\int\limits_{\bx'\in{\mathbb R}^{3}}
 \frac{1}{4\pi\vert\bx-\bx'\vert}
 {\mathbb Q}
 \left[\partial_{\tau}^{2}+\frac{1}{T^{2}}\right]
 \bJ_{1}(\bx',\tau_{\rm r})\,{\rm d}\bx'
 \\
 &-\frac{1}{c_{0}^{3}T}\int\limits_{\bx'\in{\mathbb R}^{3}}
 \frac{1}{4\pi}
 {\mathbb Q}
 \int\limits_{0}^{\tau}
 g_{1}(\bx-\bx',\tau-\tau')
 \left[\partial_{\tau'}^{2}+\frac{1}{T^{2}}\right]\bJ_{1}(\bx',\tau')\,{\rm d}\tau'\,{\rm d}\bx'
 \\
 &-\frac{1}{c_{0}^{2}}\int\limits_{\bx'\in{\mathbb R}^{3}}
 \frac{1}{4\pi\vert\bx-\bx'\vert}
 \bJ_{2}(\bx',\tau_{\rm r})\,{\rm d}\bx'
 \\
 &+\frac{1}{c_{0}^{3}T}\int\limits_{\bx'\in{\mathbb R}^{3}}
 \frac{1}{4\pi}
 \int\limits_{0}^{\tau}
 g_{1}(\bx-\bx',\tau-\tau')
 \bJ_{2}(\bx',\tau')\,{\rm d}\tau'\,{\rm d}\bx',
 \end{split} 
 \end{align}
where
 \begin{align}
 \label{eq:TauGreenFunctions}
 \begin{split}
 g_{0}(\bx,\tau)=&J_{0}\left(\frac{1}{T}\sqrt{\tau^{2}-\frac{\vert\bx\vert^{2}}{c_{0}^{2}}}\right)H\left(\tau-\frac{\vert\bx\vert}{c_{0}}\right),
 \\
 g_{1}(\bx,\tau)=&\frac{J_{1}\left(\frac{1}{T}\sqrt{\tau^{2}-\frac{\vert\bx\vert^{2}}{c_{0}^{2}}}\right)}
 {\sqrt{\tau^{2}-\frac{\vert\bx\vert^{2}}{c_{0}^{2}}}}H\left(\tau-\frac{\vert\bx\vert}{c_{0}}\right),
 \end{split}
 \end{align}
and the retarded time is again given by (\ref{eq:RetardedTau}). Notice that this retarded time appears
both explicitly in some terms in (\ref{eq:XTauFsolution}) as well as implicitly in the Green's functions as
a unit step multiplier, thus the causality in the $(\bx,\tau)$-domain is fully preserved.
Computing out the time derivatives
 \begin{align}
 \label{eq:TauDerivative}
 \begin{split}
 \left[\partial_{\tau}^{2}+\frac{1}{T^{2}}\right]\bJ_{1}(\bx,\tau)=&
 v(\tau)
 \int\limits_{0}^{\tau}u(\tau')\bJ(\bx,\tau')\,{\rm d}\tau'
 +u^{1/2}\partial_{\tau}\bJ(\bx,\tau),
 \\
 v(\tau)=&\frac{1}{T^{2}u^{1/2}}+\frac{3\partial_{\tau}u}{4u^{5/2}}-
 \frac{\partial_{\tau}^{2}u}{2u^{3/2}},
 \end{split} 
 \end{align}
we can explicitly show all the terms in (\ref{eq:XTauFsolution}) as
 \begin{align}
 \label{eq:XTauFsolutionSimplified}
 \begin{split}
 &\bE(\bx,\tau)=
 \\ 
 &\int\limits_{\bx'\in{\mathbb R}^{3}}
 \frac{3{\mathbb Q}-{\mathbb I}}{4\pi\vert\bx-\bx'\vert^{3}}
 \frac{1}{\varepsilon(\tau)}
 \int\limits_{0}^{\tau_{\rm r}}\frac{u^{1/2}(\tau)u(\tau')}{u^{1/2}(\tau_{\rm r})}\bJ(\bx',\tau')\,{\rm d}\tau'\,{\rm d}\bx'
 \\ 
 -&\int\limits_{\bx'\in{\mathbb R}^{3}}
 \frac{3{\mathbb Q}-{\mathbb I}}{4\pi\vert\bx-\bx'\vert^{2}}
 \frac{1}{c_{0}\varepsilon(\tau)}
 \int\limits_{0}^{\tau}
 g_{1}(\bx-\bx',\tau-\tau') 
 \int\limits_{0}^{\tau'}
 \frac{u^{1/2}(\tau)u(\tau'')}{Tu^{1/2}(\tau')}\bJ(\bx',\tau'')\,{\rm d}\tau''\,{\rm d}\tau'\,{\rm d}\bx'
 \\ 
 +&\int\limits_{\bx'\in{\mathbb R}^{3}}
 \frac{3{\mathbb Q}-{\mathbb I}}{4\pi\vert\bx-\bx'\vert^{2}}
 \frac{1}{c_{0}\varepsilon(\tau)}
 \int\limits_{0}^{\tau}
 g_{0}(\bx-\bx',\tau-\tau')
 \int\limits_{0}^{\tau'}
 u^{1/2}(\tau)v(\tau')u(\tau'')\bJ(\bx',\tau'')\,{\rm d}\tau''\,{\rm d}\tau'\,{\rm d}\bx'
 \\ 
 +&\int\limits_{\bx'\in{\mathbb R}^{3}}
 \frac{3{\mathbb Q}-{\mathbb I}}{4\pi\vert\bx-\bx'\vert^{2}}
 \frac{1}{c_{0}\varepsilon(\tau)}
 \int\limits_{0}^{\tau}
 g_{0}(\bx-\bx',\tau-\tau')u^{1/2}(\tau)u^{1/2}(\tau') 
 \partial_{\tau'}\bJ(\bx',\tau'')\,{\rm d}\tau'\,{\rm d}\bx'
 \\
 +&\int\limits_{\bx'\in{\mathbb R}^{3}}
 \frac{{\mathbb Q}}{4\pi\vert\bx-\bx'\vert}
 \frac{1}{c_{0}^{2}\varepsilon(\tau)}
 \int\limits_{0}^{\tau_{\rm r}}
 u^{1/2}(\tau)v(\tau_{\rm r})u(\tau')\bJ(\bx',\tau')\,{\rm d}\tau'\,{\rm d}\bx'
 \\
 -&\int\limits_{\bx'\in{\mathbb R}^{3}}
 \frac{{\mathbb Q}}{4\pi}
 \frac{1}{c_{0}^{2}\varepsilon(\tau)}
 \int\limits_{0}^{\tau}
 g_{1}(\bx-\bx',\tau-\tau') 
 \int\limits_{0}^{\tau'}
 \frac{u^{1/2}(\tau)v(\tau')u(\tau'')}{c_{0}T}\bJ(\bx',\tau'')\,{\rm d}\tau''\,{\rm d}\tau'\,{\rm d}\bx' 
 \\
 -&\int\limits_{\bx'\in{\mathbb R}^{3}}
 \frac{{\mathbb Q}}{4\pi}
 \frac{1}{c_{0}^{2}\varepsilon(\tau)}
 \int\limits_{0}^{\tau}
 g_{1}(\bx-\bx',\tau-\tau')\frac{u^{1/2}(\tau)u^{1/2}(\tau')}{c_{0}T} 
 \partial_{\tau'}\bJ(\bx',\tau')\,{\rm d}\tau'\,{\rm d}\bx' 
 \\
 +&\int\limits_{\bx'\in{\mathbb R}^{3}}
 \frac{{\mathbb Q}-{\mathbb I}}{4\pi\vert\bx-\bx'\vert}
 \frac{1}{c_{0}^{2}\varepsilon(\tau)}
 u^{1/2}(\tau)u^{1/2}(\tau_{\rm r})
 \partial_{\tau}\bJ(\bx',\tau_{\rm r})\,{\rm d}\bx'.
 \end{split} 
 \end{align}
There are some similarities here with the constant-impedance case considered in the previous section.
First of all we see both the time-weighting and the modulation of the source. However, there are also 
obvious differences. The time weighting (\ref{eq:DefTau}) contains now the square root of $a(t)$.
Hence to achive the exponential spectral shift (\ref{eq:FrequencyShift}) we need $u(t) = a_{0}+bt$,
i.e. a quadratic variation of $\varepsilon(t)$:
 \begin{align}
 \label{eq:QuadraticEps}
 \varepsilon(t)=\varepsilon_{0}(a_{0}+bt)^{2}.
 \end{align}
Further, there are new terms containing $\tau$-convolutions with the Green's functions (\ref{eq:TauGreenFunctions}).
These terms represent causal (since $\tau$-integration starts only after the wavefront has reached the 
observation point) echoes due to the scattering of the field on the temporal variation of the medium impedance.
Finally, we notice some additional longitudinal contributions in the far-field zone -- last four terms 
in (\ref{eq:XTauFsolutionSimplified}). Normally, the far-field contribution is purely transverse, i.e. contains only one
term with ${\mathbb Q}-{\mathbb I}$, see (\ref{eq:XTTimeInvariant}). Whereas, here we have three additional terms with ${\mathbb Q}$
decaying at the typical far-field $\vert\bx-\bx'\vert^{-1}$-rate. Of course, two of them are medium-induced 
echoes. However, one is definitely produced by the source alone. This means that the time-varying 
background could be used to propagate the longitudinal near-field features of the source into the far-field.

\section{Conclusions}
We considered the effects of two types of homogeneous time-varying background media on the electromagnetic field radiated
by an arbitrary source. The first type is a medium with synchronously varying permittivity and permeability, where 
the impedance stays constant at all times. Such a medium describes the time-vaying spatially flat gravitational
background, e.g. the expanding universe. The obtained analytical solution shows that under a change of
time-variable, the problem is similar to the radiation in a stationary background medium. Yet, the new time-variable
and the modulation of the source term produce significant changes in the space-time evolution of the field.
We have shown that a linear variation of the medium with time produces a ``cosmological'' frequency shift, which is an
exponential function of the distance between the source and the observer as well as an additional exponential
factor affecting the amplitude of the received signal. We have also derived the ``logarithmic'' inhomogeneous 
frequency shift for a universe expanding exponentially in time.

The more down-to-earth case of the constant permeability and time-varying pemittivity appeared, in fact, to be more difficult
and was treated using the Shwartzburg method, which involves the change of the time-variable as well as the change of the 
unknown. This method gives an analytical solution for a limited but useful class of time-varying permittivities. 
The obtained solution shows, in particular, that the medium with a quadratic temporal change in permittivity would also produce an exponential 
distance-dependent shift in the observed frequency. 
In addition, the received signal is shown to contain causal echoes induced by the scattering of waves on the time-variations 
of the impedance. The modulation of the source, which occurs in both constant and time-varying impedance cases may be 
used either for silencing or for amplification of certain sources. Surprisingly, we find that in a time-varying background, 
a DC-like constant current radiates, whereas a specific time-varying current does not.


\end{document}